\documentclass[prb,twocolumn,showpacs,superscriptaddress,floatfix]{revtex4}

\usepackage{graphicx}
\usepackage[centertags]{amsmath}
\usepackage{color}

\definecolor{grey}{rgb}{.65,.65,.65}

\newcommand{\cm}{\ensuremath{\,\mbox{cm}^{-1}}}

\newcommand{\celsius}{\ensuremath{\,{}^\circ}C}

\newcommand{\Tc}{$T_{\rm C}$}
\newcommand{\Tn}{$T_{\rm N}$}

\begin{document}

\title{Spin and lattice excitations of a BiFeO$_3$ thin film and ceramics}

\author{S.~Skiadopoulou}
\email{stella@fzu.cz}
\affiliation{Institute of Physics, The Czech Academy of Sciences, Na Slovance~2, 182
21 Prague~8, Czech Republic}
\author{V.~Goian} \affiliation{Institute of Physics, The Czech Academy of Sciences, Na
Slovance~2, 182 21 Prague~8, Czech Republic}
\author{C.~Kadlec}
\affiliation{Institute of Physics, The Czech Academy of Sciences, Na Slovance~2, 182 21 Prague~8, Czech Republic}
\author{F.~Kadlec}
\affiliation{Institute of Physics, The Czech Academy of Sciences, Na Slovance~2, 182 21 Prague~8, Czech Republic}
\author{X.F.~Bai}
\affiliation{Laboratoire SPMS, UMR 8580, CentraleSupelec, CNRS,
Universit\'e Paris-Saclay, Grande Voie des Vignes, Chatenay-Malabry, France}
\author{I.C.~Infante}
\affiliation{Laboratoire SPMS, UMR 8580, CentraleSupelec, CNRS,
Universit\'e Paris-Saclay, Grande Voie des Vignes, Chatenay-Malabry, France}
\author{B.~Dkhil}
\affiliation{Laboratoire SPMS, UMR 8580, CentraleSupelec, CNRS,
Universit\'e Paris-Saclay, Grande Voie des Vignes, Chatenay-Malabry, France}
\author{C.~Adamo} \affiliation{Department of Materials Science and
Engineering, Cornell University, Ithaca, New York, 14853 USA}
\affiliation{Department of Applied Physics, Stanford University, Stanford, CA 94305, USA}
\author{D.G.~Schlom} \affiliation{Department of Materials Science and
Engineering, Cornell University, Ithaca, New York, 14853 USA}
\affiliation{Kavli Institute at Cornell for Nanoscale Science, Ithaca, New York,
	14853 USA}
\author{S.~Kamba}
\email{kamba@fzu.cz}
\affiliation{Institute of Physics, The Czech Academy of Sciences, Na Slovance~2, 182 21 Prague~8, Czech Republic}

\date{\today}

\pacs{78.30.-j; 63.20.-e; 75.30.Ds}

\begin{abstract}

We present a comprehensive study of polar and magnetic excitations in BiFeO$_3$
ceramics and a thin film epitaxially grown on an orthorhombic (110) TbScO$_3$
substrate. Infrared reflectivity spectroscopy was performed at temperatures
from 5 to 900\,K for the ceramics and below room temperature for the thin film.
All 13 polar phonons allowed by the factor-group analysis were observed in the
ceramic samples. The thin-film spectra revealed 12 phonon modes only and an
additional weak excitation, probably of spin origin. On heating towards the
ferroelectric phase transition near 1100\,K, some phonons soften, leading to an
increase in the static permittivity. In the ceramics, terahertz transmission
spectra show five low-energy magnetic excitations including two which were not
previously known to be infrared active; at 5\,K, their frequencies are
53 and 56\cm. Heating induces softening of all magnetic modes. At a temperature
of 5\,K, applying an external magnetic field of up to 7\,T irreversibly alters
the intensities of some of these modes. The frequencies of the observed spin
excitations provide support for the recently developed complex model of
magnetic interactions in BiFeO$_3$ (R.S.\ Fishman, Phys.\ Rev.\ B \textbf{87},
224419 (2013)). The simultaneous infrared and Raman activity of the spin excitations is consistent with their assignment to electromagnons.
\end{abstract}

\maketitle

\section{Introduction}

Among novel materials, an intense effort is concentrated on the
study of multiferroics.\cite{Fiebig05,Tokura14-review} A wide range of
applications, such as information storage, sensing, actuation, and spintronics,
await pioneering materials and strategies that would produce robust
magnetoelectric coupling at room temperature (RT).\cite{Eerenstein06, Spaldin10}
The  ability to manipulate
magnetization in a magnetoelectric multiferroic by
electric fields can be extremely promising for such applications, due to the
simplicity and cost-efficiency of applying an electric field. As one of the few single-phase RT magnetoelectric
multiferroics, bismuth ferrite BiFeO$_3$, is at the center of  attention, as it presents a ferroelectric
phase transition at approximately 1100\,K and an antiferromagnetic one at
643\,K.\cite{Catalan09}

The knowledge of lattice and spin excitations in BiFeO$_3$ is essential for
understanding the underlying mechanisms that induce its multiferroic
behavior. A series of Raman and Infrared (IR) spectroscopy studies have
presented controversial results concerning the assignment of the magnon and
phonon modes, as well as of the highly acclaimed electromagnons
(i.e., electrically active magnons). Probing IR-active low-energy excitations is hindered by a lack of sufficiently large single crystals.
 Raman-active phonons of the
rhombohedral $R3c$ BiFeO$_3$ structure have been reported for single
crystals,\cite{Haumont06, Fukumura07, Palai10b} polycrystalline
ceramics,\cite{Kothari08, Rout09, Hlinka11} and thin films,\cite{Singh06,
Palai10} however, there is a significant discrepancy between the claimed
phonon frequencies and symmetry representations.  A possible explanation for
such inconsistences between the various experimental results is the
presence of oblique phonon modes, which show a continuous variation of frequency
along the phonon propagation vector with respect to the crystallographic axis of
the probed specimen.\cite{Hlinka11} Up to now, the phonon IR spectroscopy
studies have been focused on ceramics\cite{Kamba07, Komandin10, Massa10} and
single crystals,\cite{Lobo07, Lu10} whereas, to our knowledge, no report
of a thin-film IR investigation exists.

When it comes to magnon and electromagnon studies, Raman spectroscopy
holds the record for the number of excitations observed, yielding up to 16
magnetic modes corresponding to the cycloidal spin structure (8 $\Phi_n$ cyclon
and 8 $\Psi_n$ extra-cyclon modes) for single crystals,\cite{Singh08,
Cazayous08, Rovillain09, Rovillain10} but only one for polycrystalline thin
films.\cite{Kumar11, Kumar12} Furthermore, the influence of strain on the
number of spin excitations and their frequencies were reported for epitaxial
thin films grown on a series of substrates.\cite{Sando13} Time-domain THz
spectroscopy,\cite{Talbayev11} inelastic neutron scattering
measurements\cite{Jeong12, Delaire12, Xu12, Jeong14} and absorbance spectroscopy
in the THz range\cite{Nagel13} have revealed modes similar to the Raman-active
onespredicted by theoretical calculations.\cite{Nagel13,deSousa08,
Fishman12, Fishman13a, Fishman13b} The best agreement between the experimental
and theoretical spin-excitation frequencies (including their splitting in
external magnetic field) was obtained for a microscopic model that takes into
account the nearest and next-nearest neighbor exchange interactions, two
Dzyaloshinskii-Moriya interactions and an easy-axis anisotropy.\cite{Nagel13,
Fishman13b} The same model successfully described the low-energy inelastic
neutron scattering spectra.\cite{Jeong14}  In contrast, Komandin \textit{et
al.}\cite{Komandin10} showed IR transmission spectra with an excitation at
approximately 47\cm{} which had not been previously reported by experimental or
theoretical studies. The intensive discussion in the
literature concerning the nature of the spin excitations raises the question:
which excitations are pure magnons (i.e., contribute only to the magnetic
permeability $\mu$) and which are electromagnons (i.e., influence at least
partially the  permittivity $\varepsilon^*$)?  It is worth noting that according
to a recent symmetry analysis,\cite{Szaller14} BiFeO$_3$ allows directional
dichroism and therefore spin waves can be simultaneously excited by the
electric and magnetic components of electromagnetic radiation. For more details
on BiFeO$_3$ spin dynamics, see the review of Park \textit{et al}.\cite{Park14}

In the current work, we report spin and lattice excitations in BiFeO$_3$
ceramics, as measured by the combination of IR reflectivity and time-domain THz
transmission spectroscopy, in a temperature range from 10 to 900\,K. All 13
IR-active phonon modes are observed, exhibiting softening on
heating. Five low-frequency spin modes are detected from 5\,K up
to RT,
the highest two appearing at 53 and 56\cm. This corresponds to the frequency
range where such excitations were
theoretically predicted,\cite{Nagel13,deSousa08,Fishman13b} but not experimentally
confirmed up to now. At 5\,K, the low-energy spin dynamics
in the THz range were also studied in a
varying magnetic field of up to 7\,T. Softening of the (electro)magnon
frequencies upon increasing the magnetic field was observed. Additionally,
a BiFeO$_3$ epitaxial thin film grown on an orthorhombic (110) TbScO$_3$
single crystal substrate
was studied for the first time via IR reflectance spectroscopy.

\section{Experimental Details}

BiFeO$_3$ ceramics were prepared by the solid-state route. A
stoichiometric mixture of Fe$_2$O$_3$ and Bi$_2$O$_3$ powder
oxides with a purity of 99.99\% was ground and uniaxially cold-pressed under 20-30\,MPa pressure
into 8\,mm diameter pellets.  The pellets were then covered by sacrificial BFO
powder to avoid bismuth oxide loss from the pellet,
and sintered in a tube
	furnace at 825\celsius{} for8\,h in air. To avoid any
secondary phase formation, the samples were quenched to RT.
Polished disks with a diameter of 4\,mm and thicknesses of ca.\ 600 and
338\,$\mu$m were used for the IR reflectivity and THz transmission
measurements, respectively.

 An epitaxial BiFeO$_3$ thin film with a thickness of
300\,nm was
 grown by reactive molecular-beam epitaxy on a (110) single crystal
 substrate. The growth parameters were the same as for the samples reported in
 Ref.~\onlinecite{Nelson11}. Near-normal incidence IR reflectivity spectra
of the BiFeO$_3$ ceramics and film were obtained using a
Fourier-transform IR spectrometer Bruker IFS 113v in the frequency range
20--3000\cm{} (0.6--90\,THz) at RT; for the low and high temperature
measurements the spectral range was reduced  to 650\cm{}. Pyroelectric
deuterated triglycine sulfate detectors were used for the room- and
high-temperature measurements up to 900\,K, whereas a He-cooled (operating
temperature 1.6\,K) Si bolometer was used for the low-temperature measurements
down to 10 K. A commercial high-temperature cell (SPECAC P/N 5850) was used for
the high-temperature experiments. The thermal radiation from the hot
sample entering the interferometer was taken into account in our spectra
evaluation.

THz measurements from 3\cm{} to 60\cm{} (0.09--1.8\,THz) were
performed in the transmission mode with the use of a custom-made time-domain
terahertz spectrometer. In this spectrometer, a femtosecond Ti:sapphire laser
oscillator (Coherent, Mira) produces a train of femtosecond pulses which
generates linearly polarized broadband THz pulses in a photoconducting switch
TeraSED (Giga-Optics). A gated detection scheme based on electrooptic sampling
with a 1 mm thick [110] ZnTe crystal as a sensor allows us to measure the time
profile of the electric field of the transmitted THz pulse. The same
high-temperature cell as for the IR reflectivity was used for the THz range
high-temperature measurements. Oxford Instruments Optistat
optical cryostats with mylar and polyethylene windows were used for
the low-temperature THz and IR measurements, respectively. THz experiments in
an external magnetic field $H_{\rm ext}\leq7$\,T were performed upon
decreasing $H$ with an Oxford
Instruments Spectromag cryostat in the Voigt configuration, where the electric
component of the THz radiation \textbf{E}$_{\rm THz}$ was set perpendicular to
$H_{\rm ext}$.

The IR reflectivity and THz complex (relative) permittivity spectra
$\varepsilon^{\ast}(\omega)$ were carefully fit assuming the factorized
form of the dielectric function based on a generalized
damped-harmonic-oscillator model:\cite{Gervais83}
\begin{equation}
\label{eq:1}\varepsilon^{\ast}(\omega)=\varepsilon_\infty\prod_{j=1}^N\frac{\omega^2_{\rm LOj}-\omega^2+i\omega\gamma_{\rm LOj}}{\omega^2_{\rm TOj}-\omega^2+i\omega\gamma_{\rm TOj}} ,\end{equation}
where $\omega_{\rm TOj}$ and $\omega_{\rm LOj}$ are the frequencies of the $j$-th
transverse optical (TO) and longitudinal optical (LO) phonons,
$\gamma_{\rm TOj}$ and $\gamma_{\rm LOj}$ are the corresponding damping constants, and
$\varepsilon_\infty$ denotes the high-frequency (electronic) contribution to the
permittivity, determined from the RT frequency-independent reflectivity tail above the phonon frequencies. The reflectivity $R(\omega)$ is related to the complex dielectric function  $\varepsilon^{\ast}(\omega)$ by:
\begin{equation}\label{eq:2}R(\omega)=\left|\frac{\sqrt{\varepsilon^{\ast}(\omega)}-1}{\sqrt{\varepsilon^{\ast}(\omega)}+1}\right|\,.
\end{equation}

To evaluate the IR reflectance spectra of the BiFeO$_3$/TbScO$_3$  thin
film, a model corresponding to a two-layer optical system was
used.\cite{Zelezny98} The IR reflectivity spectra of the bare TbScO$_3$
substrate were fit first at each temperature, using the model described
by Eq.~\ref{eq:1}. The values of oscillator parameters
($\omega_{\rm TOj}$, $\omega_{\rm LOj}$, $\gamma_{\rm TOj}$,  $\gamma_{\rm
LOj}$, and
$\varepsilon_\infty$) obtained for the substrate were used for
 fitting the BiFeO$_3$/TbScO$_3$  thin-film
spectra. The model used to describe the complex permittivity of the thin
film is a sum of $N$ independent three-parameter damped harmonic
oscillators, which can be expressed as:\cite{Gervais83}
\begin{equation}
\label{eq:3}\varepsilon^{\ast}(\omega)=\varepsilon_\infty+\sum_{j=1}^N\frac{\Delta\varepsilon_j\omega^2_{\rm
TOj}}{\omega^2_{\rm TOj}-\omega^2+i\omega\gamma_{\rm TOj}},
\end{equation}
where $\Delta\varepsilon_j$ is the dielectric strength of the $j$-th mode.
Eq.~\ref{eq:3} is simpler than Eq.~\ref{eq:1}, and its use is well
justified because the damping of the LO phonons of the film does not
appreciably influence the reflectance spectra.

\section{Results and Discussion}

\subsection{Phonons in the BiFeO$_3$ ceramics}

\begin{figure}
\begin{center}
\includegraphics[width=0.9\columnwidth]{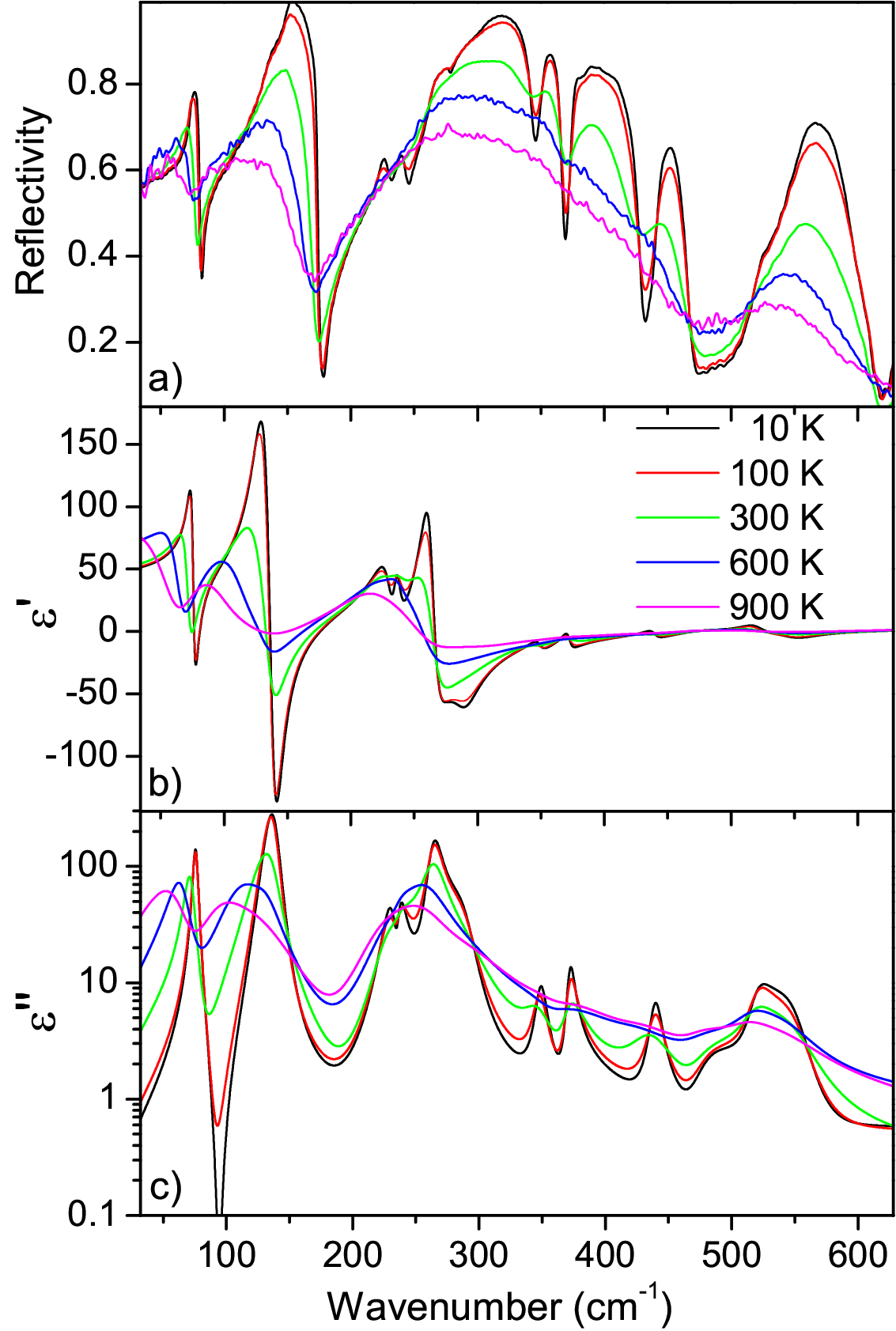}
\end{center}
\caption{(Color online) (a) IR reflectivity spectra of BiFeO$_3$ ceramics at selected temperatures; (b) Real, and (c) imaginary part of permittivity, as obtained from the fits.}
\label{fig1}
\end{figure}

The temperature dependence of the experimental IR reflectivity spectra of
the BiFeO$_3$ ceramics is shown in Fig.~\ref{fig1}(a). The reflectivity band
intensities are higher than those in the previously published spectra of
ceramics\cite{Kamba07, Komandin10, Massa10} and single crystal,\cite{Lu10} but
comparable to the single-crystal spectra published by Lobo
\textit{et al}.\cite{Lobo07} This confirms the very high quality and
density of our ceramic samples, which is essential for an accurate
determination of phonon and magnon parameters.

\begin{figure}
\begin{center}
\includegraphics[width=0.95\columnwidth]{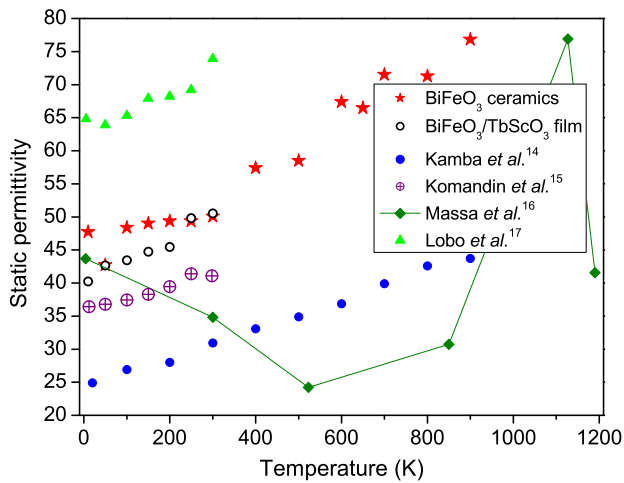}
\end{center}
\caption{(Color online) Temperature dependence of the static permittivity calculated
	from the sum of phonon contributions. The data
	for the ceramics and the film (\textbf{E}$_{\rm ext}
	\parallel[1\bar{1}0]$) from the current work are compared with
previously published data obtained, in order, on ceramics,\cite{Kamba07, Komandin10,
Massa10} and a single crystal.\cite{Lobo07} The line connecting the data
from Ref.~\onlinecite{Massa10} is a guide for the eyes. See the
text for additional details.}
\label{fig2}
\end{figure}

The IR reflectivity and THz transmission spectra were fit
simultaneously using Eq.~\ref{eq:1}; the
resulting phonon parameters are listed in Table~1. The real and imaginary
parts of the complex dielectric spectra calculated from the fits are shown in
Figs.~\ref{fig1}(b),(c). As predicted by the factor group analysis for the
rhombohedral $R3c$ structure  of BiFeO$_3$,\cite{Kamba07} all 13 IR-active
phonons (4$A_1$ + 9$E$ symmetries) are clearly seen from 10\,K up to RT. In
Table~1, the symmetries of all modes are assigned based on Raman
spectra,\cite{Hlinka11} first-principles calculations,\cite{Hermet07} and IR
spectra of single crystals.\cite{Lobo07} The damping of the modes strongly
increases on heating; therefore, above 300\,K, the IR reflection bands
broaden and mutually overlap. Note the remarkable softening of the modes below
150\cm{} upon heating (see Fig.~\ref{fig1}) . Consequently, the static permittivity
$\varepsilon$(0) determined by the sum of phonon contributions as
$\varepsilon(0)=\varepsilon_{\infty}+\sum{\Delta\varepsilon_{j}}$ increases on
heating in agreement with the Lyddane-Sachs-Teller
relation\cite{Lyddane41} (see Fig.~\ref{fig2}). A qualitatively similar
temperature dependence was published in previous studies of BiFeO$_3$
ceramics,\cite{Kamba07, Komandin10} but the absolute value of permittivity was
lower, probably due to a lower density/quality of the earlier
studied ceramics. In contrast, the permittivity calculated from the
single-crystal spectra\cite{Lobo07} is slightly higher, due to the absence of the $A_1$ modes
whose dielectric strength $\Delta\varepsilon_j$ is lower than the $E$-symmetry
modes.

\begin{table*}
\caption{A comparison of the  parameters of the IR-active modes in the BiFeO$_3$ ceramics (at
10 and 900 K) and the thin film (at 10 K), obtained from  IR
spectra fitting. The symmetry assignment of each mode is also given.}

\begin{ruledtabular}
\begin{tabular}{crrrrrrrrrrrrrrr}
  & \multicolumn{8}{c}{Ceramics} & \multicolumn{6}{c}{Thin film} \\
\cline{2-9} \cline{10-15}
  & \multicolumn{4}{c}{10 K} & \multicolumn{4}{c}{900 K} & \multicolumn{3}{c}{10 K (\textbf{E}$_{\rm ext} \parallel[1\bar{1}0]$)} & \multicolumn{3}{c}{10 K (\textbf{E}$_{\rm ext} \parallel[001]$)} \\
\cline{2-5} \cline{6-9} \cline{10-12} \cline{13-15}
   Symmetry & $\omega_{\rm TO}$ & $\gamma_{\rm TO}$ & $\omega_{\rm LO}$ & $\gamma_{\rm LO}$ & $\omega_{\rm TO}$ & $\gamma_{\rm TO}$ & $\omega_{\rm LO}$ & $\gamma_{\rm LO}$ & $\omega_{\rm TO}$ & $\gamma_{\rm TO}$ & $\Delta\varepsilon$ & $\omega_{\rm TO}$ & $\gamma_{\rm TO}$ & $\Delta\varepsilon$\\
\hline

\hline
 $E$ & 75.8 & 4.6 & 81.5 & 2.8 & 64.3 & 40.9 & 70.5 & 19.3 & 74.7 & 1.4 & 4.3 & 77.8 & 1.8 & 4.5 \\
 $E$ & 133.7 & 22.8 & 137.4 & 34.6 & 101.0 & 65.1 & 117.9 & 51.1 & 132.2 & 2.3 & 8.6 & 136.7 & 2.9 & 12.2 \\
 $A_1$ & 137.8 & 13.6 & 175.8 & 4.0 & 124.3 & 57.6 & 165.8 & 32.9 & 145.2 & 2.4 & 7.4 \\
 $A_1$ & 231.2 & 14.0 & 233.9 & 5.6 & 226.2 & 59.6 & 233.5 & 31.4 & 225.9 & 2.9 & 0.6 & 227.4 & 2.2 & 0.7 \\
 $E$ & 236.7 & 7.6 & 243.6 & 13.3 & 234.0 & 31.7 & 240.8 & 69.3 & 241.0 & 4.3 & 0.9 & 242.4 & 4 & 0.2 \\
 $E$ & 264.0 & 13.0 & 283.0 & 38.0 & 258.2 & 60.4 & 278.0 & 48.5 & 266.2 & 2.3 & 5.4 & 266.8 & 4.4 & 0.4 \\
 $E$ & 288.5 & 26.2 & 345.8 & 9.2 & 278.4 & 61.6 & 340.7 & 40.4 & 293.9 & 4.6 & 3.6 & 279.5 & 5.5 & 10.6 \\
 $E$ & 349.0 & 10.0 & 368.1 & 5.0 & 342.5 & 42.5 & 363.6 & 41.0 & & & \\
 $E$ & 372.0 & 7.3 & 430.7 & 10.6 & 364.5 & 50.1 & 420.3 & 51.4 & & & & 371.7 & 9.2 & 0.9\\
 $E$ & 439.6 & 11.0 & 468.0 & 13.3 & 424.0 & 54.3 & 464.3 & 37.1 & & & & 445.7 & 0.3 & 1.1 \\
 $A_1$ & 476.2 & 41.2 & 503.7 & 49.1 & 469.8 & 41.0 & 502.8 & 106.2 & 472.2 & 7.1 & 0.01 & \\
 $E$ & 520.2 & 21.8 & 523.8 & 76.0 & 514.7 & 45.7 & 518.3 & 62.5 & 524.0 & 9.8 & 0.3 & 516.8 & 12.7 & 0.04 \\
 $A_1$ & 549.5 & 36.9 & 606.2 & 27.1 & 542.1 & 78.8 & 595.3 & 102.3 & 550.6 & 7.2 & 0.07 & 551.8 & 2.3 & 1 \\
 spin wave &  &  &  &  &  &  &  &  &  &  &  & 603.6 & 9.9 & 0.01 \\
            \end{tabular}
 \end{ruledtabular}
\end{table*}

We would like to stress that the temperature dependence of permittivity
calculated from the phonon contributions has been previously published mostly below
RT,\cite{Lobo07, Komandin10} and only Ref.~\onlinecite{Kamba07} reported
$\varepsilon(0,T)$ up to 900\,K. Experimental low-frequency (i.e., below 1\,MHz)
dielectric data are also not available above RT due to the presence of
significant
leakage currents in BiFeO$_3$ at high temperatures. There is only one
publication\cite{Massa10}  presenting IR spectra of BiFeO$_3$ ceramics up to 1280\,K,
i.e., far above \Tc, however, in this case, emissivity was measured instead of
reflectivity above 600\,K and
$\varepsilon(0,T)$ was not reported. In that work, Massa \textit{et
al.}\cite{Massa10} listed phonon parameters at selected temperatures, from which
we re-calculated $\varepsilon(0,T)$; these values are presented for sake of
comparison in Fig.~\ref{fig2}. One can see that, at $T=523$ and 850\,K,
Massa \textit{et al.}\cite{Massa10} obtained the lowest published values of $\varepsilon(0)$. We also note the unusual increase in $\varepsilon_{\infty}$ and $\varepsilon(0,T)$ on cooling below RT; the latter appears to be in contradiction with the soft-mode frequency increase on cooling, reported in the same work.
Finally, two additional phonon modes at $T=4$\,K, not
predicted by the factor-group analysis, are reported in Ref.\
	\onlinecite{Massa10}. All these facts raise questions about
	whether these results describe the intrinsic properties of
	BiFeO$_3$.  Nevertheless, it is worth noting the peak in
Massa's data at the ferroelectric phase transition near 1120\,K; this is
the only experiment showing a dielectric anomaly at $T_{\rm C}$ in BiFeO$_3$.
The peak is much lower than in other canonical ferroelectrics, apparently
due to the first-order character of the phase transition from
$R3c$ symmetry to the high-temperature $Pnma$
structure.\cite{Catalan09, Park14}

\subsection{Excitations in the BiFeO$_3$ thin film}

\begin{figure}
  \begin{center}
   \includegraphics[width=0.90\columnwidth]{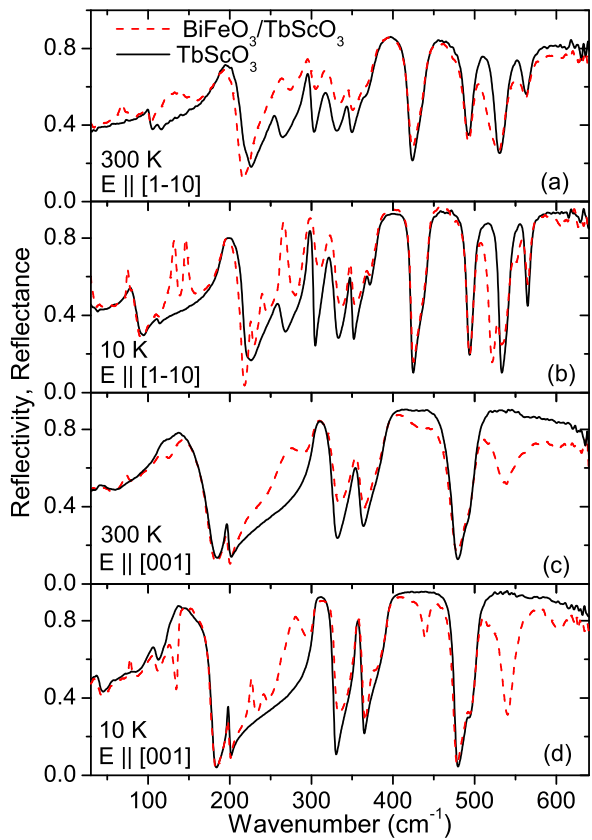}
  \end{center}
    \caption{(Color online) Room-temperature and 10\,K IR reflectivity spectra of
    the TbScO$_3$  substrate and reflectance of the BiFeO$_3$/TbScO$_3$  thin film for
    polarizations (a,b) \textbf{E}$_{\rm ext} \parallel[1\bar{1}0]$ and (c,d) \textbf{E}$_{\rm ext} \parallel[001]$.}
    \label{fig3}
\end{figure}

The  polarized IR reflectivity spectra of the bare (110) TbScO$_3$  substrate and
reflectance spectra of the BiFeO$_3$/TbScO$_3$  thin film are shown in Fig.~\ref{fig3}.
All BiFeO$_3$ mode frequencies obtained by fitting are listed in Table~1. We note that, on the one hand, the BiFeO$_3$ phonons
below 150\,\cm{} are better resolved in the \textbf{E}$_{\rm ext}
\parallel[1\bar{1}0]$ polarization with respect to the substrate crystal axes,
because the TbScO$_3$  phonons are weak in this case. On the other hand, the BiFeO$_3$ modes
at 372 and 446\,\cm{} are seen only in the \textbf{E}$_{\rm ext}
\parallel[001]$ polarized spectra, due to a more favorable TbScO$_3$  spectrum in this
region. Namely, the large TO-LO splitting of the TbScO$_3$  phonons within
370--480\,\cm{} enhances the sensitivity of the IR reflectance to the thin-film
phonons.\cite{Goian14} Therefore, due to the properties of TbScO$_3$ , the two polarizations provide complementary
information on the BiFeO$_3$ response. Note that
some TbScO$_3$  phonons seen below 150\cm\, (see Fig.~\ref{fig3}) exhibit anomalous
temperature shifts and even splitting  on cooling. This is probably caused by
phonon interaction with the crystal field (and related electronic
transitions).\cite{Goian13}

The in-plane lattice parameters of the (110) TbScO$_3$  substrate are slightly smaller
than those of BiFeO$_3$, inducing only a small compressive strain ($\sim0.24\%$) in
the (001)$_{\rm cub}$ epitaxial film. Here, the orientation of the thin film is marked
with respect to the pseudocubic crystal axis (which is denoted by the
subscript ``cub'') of BiFeO$_3$. The ferroelastic and
ferroelectric domain structure of BiFeO$_3$/TbScO$_3$  was investigated in
Refs.~\onlinecite{Folkman09,Nelson11} where two kinds of stripe-like domains separated by
(010)$_{\rm cub}$ vertical boundaries were reported. The spontaneous
polarization \textbf{P} in
adjacent domains is rotated by 109$^{\circ}$ and its direction is tilted from
the [001]$_{cub}$ direction.\cite{Folkman09}
In our IR spectra, only vibrations polarized in the film plane are active. Since the
ferroelectric polarization \textbf{P} is tilted from the normal of the thin film
plane, phonons of both $E$ and $A_1$ symmetries can be excited. Nevertheless, we see only
some of the modes due to our limited sensitivity to the thin film.
In contrast to the ceramics samples, the phonon damping observed in the
thin film is much lower (see Table~1). The values of the damping
constants are comparable to those of single
crystals,\cite{Lobo07} which confirms the high quality of
our epitaxial thin film. Similar to the case of BiFeO$_3$ ceramics,
the phonon eigenfrequencies in the film decrease on heating, leading to an increase
in the static permittivity $\varepsilon(0,T)$ (see
Fig.~\ref{fig2}). The phonon frequencies of the BiFeO$_3$ ceramics and the
film present no significant differences, which is clearly a
consequence of the very
small strain applied to the film by the substrate. The
calculated $\varepsilon(0,T)$ is, however, smaller in the BiFeO$_3$ film than in
the single crystal (see Fig.~\ref{fig2}), because, for the
former sample, only strong phonons were revealed in the IR reflectance
spectra.

In the low-temperature \textbf{E}$_{\rm ext} \parallel[001]$ spectra, a weak but
sharp and clearly observable minimum in reflectance develops near
600\,\cm{} (see Fig.~\ref{fig3}(d)). We can exclude its phonon origin, because the
factor-group analysis in rhombohedral structure does not allow an additional mode. A lower crystal symmetry is also excluded because our thin film has only 0.24\% compressive strain; it is known that the films change the structure only with strain higher than 2\%.\cite{Sando14} Also the multi-phonon
origin is not likely since multi-phonon scattering usually decays on cooling; at
RT, the peak intensity is markedly lower, close to our sensitivity limit (see
Fig.~\ref{fig3}(c)). Interestingly, the position of the peak, ca.\ 604\,\cm,
corresponds to the maximum magnon energy observed out of the Brillouin-zone
center by inelastic neutron
scattering in a BiFeO$_3$ single crystal.\cite{Jeong12} Therefore, we assign this peak
to scattering by the highest-energy part of the magnon branch. The magnon
density of states can be activated in the IR spectra due to the modulated cycloidal magnetic
structure of BiFeO$_3$; the maxima in the density of states occur either at the
maximal energy or at the Brillouin-zone boundary. Correspondingly, another peak
in the density of states can be expected for the magnon branch at the $K$
and $M$ points.
These magnons have an energy\cite{Jeong14} of ca.\ 525\,\cm, which falls into
the range of the highest-energy phonon. We assume that this strong phonon
screens the weak
magnon signal, which is why we do not detect the corresponding absorption in the
IR spectra. Finally, let us mention that the magnons from the Brillouin-zone boundary become
frequently electrically active due to the exchange striction;\cite{Stenberg12} therefore, they
can be called electromagnons.

\subsection{Magnetic excitations in THz spectra of BiFeO$_3$ ceramics}
\begin{figure}
\begin{center}
\includegraphics[width=0.8\columnwidth]{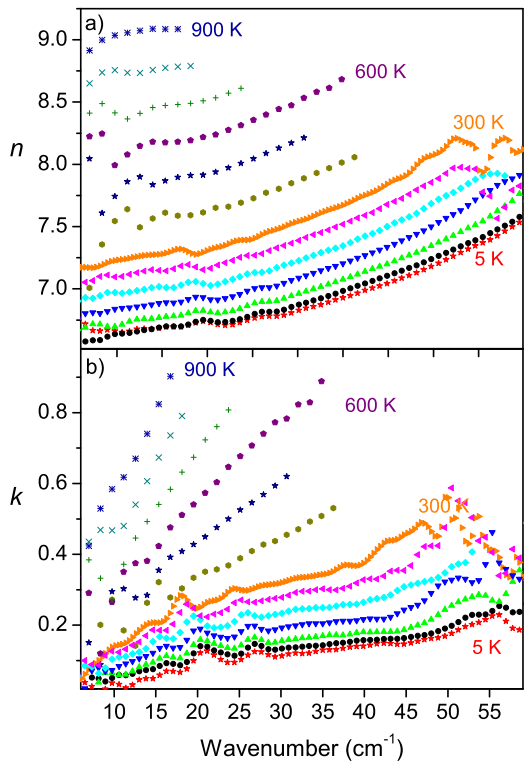}
\end{center}
\caption{(Color online) Temperature dependences of the  (a)
	refractive index and (b) extinction coefficient spectra of
	BiFeO$_3$ ceramics, determined from the THz transmission.
The spectra are shown for every 50 and 100\,K below and above RT, respectively.}
\label{fig4}
\end{figure}

Our efforts to measure THz spectra of the BiFeO$_3$ film failed, due to
insufficient signal. This can be explained by the fact that the thin
film only slightly absorbs the THz pulses and the index of
refraction $n$ of BiFeO$_3$ is higher than that of the substrate only by $\approx$ 50\%. We
note that time-domain THz spectroscopy can be used only for studies of the thin
films with $n$ at least one order of magnitude higher than that of the
substrate. As an example, such a technique was successfully used for investigation of the ferroelectric soft mode behavior near a strain-induced ferroelectric phase transition in SrTiO$_3$ thin films.\cite{Nuzhnyy09}

In contrast, THz spectra of the BiFeO$_3$ ceramics were successfully measured at
temperatures from 5 to 900\,K. The spectra of the complex
refractive index $N(\omega)=n(\omega)+ik(\omega)$, determined from
experimental data for various temperatures,
are presented in Fig.~\ref{fig4}. In magnetic systems, $N$ depends on
the complex  permittivity $\varepsilon^*$ and  permeability
$\mu^*$ via the relationship $N(\omega)=\sqrt{\mu^*\varepsilon^*}$.  Since we cannot resolve
 whether the observed modes contribute
to $\mu^*$ (as magnons) or $\varepsilon^*$ (as electromagnons or polar phonons), we
only present $n(\omega)$, $k(\omega)$ spectra in Fig.~\ref{fig4}.

\begin{figure}
	\begin{center} \includegraphics[width=0.85\columnwidth]{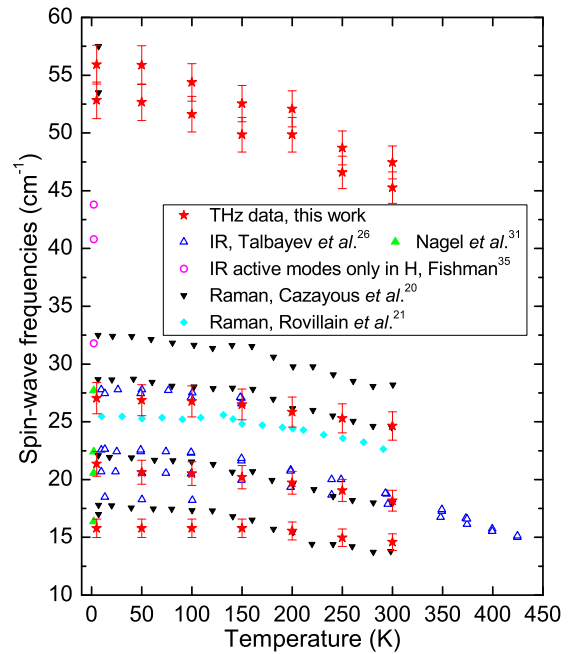}
	\end{center} \caption{(Color online) Temperature dependence of the
	spin-excitation frequencies determined by fits of the absorption index spectra in the THz range compared with published frequencies obtained from Raman scattering\cite{Cazayous08,Rovillain09} and far IR spectra\cite{Nagel13}. The modes at $\sim$ 32, 41 and 44 \cm\, were predicted by  Fishman\cite{Fishman13b} and they become IR--active only in magnetic field above 7\,T\cite{Nagel13}.} \label{fig5} \end{figure}

\begin{figure} \begin{center} \includegraphics[width=0.85\columnwidth]{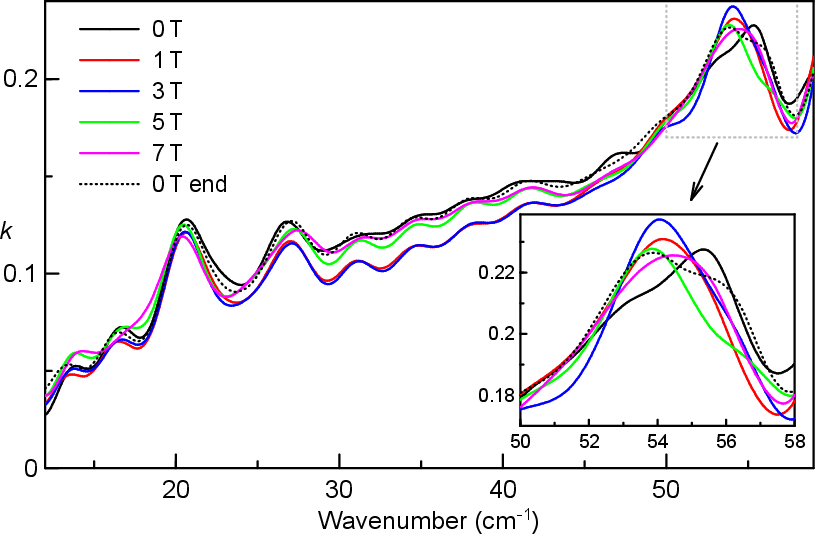}
\end{center} \caption{(Color online) Magnetic field dependence of the
	experimental THz extinction coefficient in the BiFeO$_3$ ceramics
	measured at 5\,K. The dotted line shows the zero-field spectrum
	after applying the magnetic field.} \label{fig6}
\end{figure}
	
In the THz spectra, one can see a gradual increase in $n$ and $k$ on
heating. This is mainly caused by the softening of the TO$_1$ phonon and
an increase in the phonon damping with heating.  The weak maxima in the
$k(\omega)$ spectra correspond to frequencies of magnetic excitations.
Their temperature dependence is plotted in Fig.~\ref{fig5} together with the modes observed in Raman\cite{Cazayous08,Rovillain09} and IR\cite{Talbayev11,Nagel13} spectra. Note that the modes between 32 and 44 \cm\ become IR--active only in external magnetic field.\cite{Nagel13,Fishman13b}  Besides a good agreement between the frequencies of Raman and IR--active modes, one can
see a gradual decrease in all mode frequencies with increasing temperature.
Although the excitations should exist at least up to \Tn,  we fit our THz spectra
only up to 300\,K, because, above RT, their damping is very high,
precluding their exact fitting. We clearly distinguish five
magnetic excitations. The three of them appearing up to 27\cm{}
correspond to the IR-active modes observed earlier by other
authors.\cite{Komandin10, Talbayev11, Nagel13} In our spectra, they
exhibit an enhanced damping in comparison with single-crystal data,
mainly due to the fact that we measured un-polarized spectra. At 5\,K,
two modes are seen with peak absorption frequencies of	$53$\cm{} and
$56$\cm{}. These modes were not observed in IR
spectra before, apparently because the samples used in previous studies
were opaque above 40\cm{}. These  peak frequencies correspond well
to the $\Psi_7$ and $\Phi_8$ modes reported in Raman spectra
by Cazayous \textit{et al.}.\cite{Cazayous08} Note that Komandin
\textit{et al.}\cite{Komandin10} predicted a polar and
heavily damped mode at an estimated frequency of 47\cm, but their rough
estimation was based on discrepancies between IR reflectivity and THz
transmission spectra. Nagel \textit{et al.}\cite{Nagel13} discovered
additional spin excitations near 40 and 43\cm, but a magnetic field higher
than 5\,T was needed for the activation of these excitations in
far-IR transmission
spectra. Furthermore, Fishman predicted\cite{Fishman13b} a mode above
50\cm, which should soften in an external magnetic field,
and de Sousa \textit{et al.} predicted\cite{deSousa08} exactly the two modes which we observe.

We measured THz transmittance at 5\,K under a magnetic field up to 7\,T (see
the $k(\omega)$ spectra in Fig.~\ref{fig6}). All modes exhibit small
frequency shifts
with increasing magnetic field. A similar behavior was reported by Nagel
\textit{et al.},\cite{Nagel13} but their frequency shifts were higher
due to much higher values of the applied magnetic field (up to 31\,T).
We see also weak indications of the modes near 32, 35, 38, and 42\cm,
which correspond well (except the one at 35\cm) to the magnetic-field
induced modes of Ref.~\onlinecite{Nagel13}. In our measurements, while the sample was
kept at a temperature of 5\,K, applying the magnetic field of 7\,T
irreversibly changed the shape of the magnetic modes in the absorption index spectrum
(see Fig.~\ref{fig6}), similarly to the observations in
Ref.~\onlinecite{Nagel13}. This behavior is probably a consequence of
magnetic-field-induced changes in the geometry of magnetic domains and pinning
the domain walls on defects. The  pronounced modes near
53 and 56\,\cm{}  can be identified with the pair of modes predicted by
Fishman near 45\,\cm{} at $H_{\rm ext}=15$\,T.\cite{Fishman13b} Thus, our observations confirm the complex
effective Hamiltonian describing the magnetic interactions in BiFeO$_3$; it
includes nearest and next-nearest-neighbor exchange interactions, two
Dzyaloshinskii-Moriya interactions, and an easy-axis
anisotropy.\cite{Fishman13b} Very recently, this model was also
confirmed by low-energy inelastic neutron scattering spectra.\cite{Jeong14}

We were not able to fit unambiguously the THz spectra above 300\,K using the
modes seen at low temperatures, owing to their heavy damping at high
temperatures. Nevertheless, a fit with one effective overdamped mode
provided
reasonable results. Surprisingly, this overdamped mode with a relaxation frequency
around 15\cm{} and $\Delta\varepsilon=0.5$--1.2 ($\Delta\varepsilon$ rising with
T) was necessary even above \Tn, suggesting that there are still some paramagnons
present in the paramagnetic phase. Alternatively, this mode might be due to a multi-phonon or
quasi-Debye absorption allowed in the non-centrosymmetric
phase.\cite{Tagantsev03}

Finally, we would like to stress that from unpolarized THz spectra, we could not
distinguish whether our magnetic excitations are magnons or electromagnons.
 The symmetry analysis published by Szaller \textit{et
al.},\cite{Szaller14} however, allows electromagnons in the cycloidal G-type
antiferromagnetic phase of BiFeO$_3$.  In analogy with selection rules for polar
phonons which, in acentric lattices, are both IR and Raman active, we suggest
that for spin excitations, their simultaneous IR and Raman activites
indicate that they are electromagnons. During the review process, the polar activity of spin excitations has been confirmed by K\'{e}zsm\'{a}rki \textit{et al.},\cite{Kezsmarki15} who observed directional dichroism\cite{Takahashi12, Szaller14} of a BiFeO$_3$ single crystal in polarized far-IR spectra between 10 and 30\cm. 

\section{Conclusions}
An  extensive study of IR vibrational
spectra of BiFeO$_3$ ceramics and an epitaxial thin film is
reported. The intensities of all phonons observed in the ceramics
are higher than in the previous publications.\cite{Kamba07,
Komandin10, Massa10} Thus, the static permittivity of our samples
calculated from phonon contributions is close to the previously published
single crystal data.\cite{Lobo07} Some phonons slightly soften on heating,
leading to an increase in the permittivity towards \Tc.
Nevertheless, the permittivity is much lower than in canonical ferroelectric
perovskites, such as BaTiO$_3$ and KNbO$_3$, because the  phonons in BiFeO$_3$ are much harder
and the Born effective charges are much smaller.
In addition, the phonons in a BiFeO$_3$/TbScO$_3$
epitaxial thin film were studied for the first time, showing
parameters similar to those in BiFeO$_3$ single crystals. In the thin film, an
additional weak excitation near 600\,\cm{} was detected which apparently
corresponds to a peak in the
magnon density of states and which can be excited due to
the incommensurately modulated magnetic structure of BiFeO$_3$. Time-domain THz
spectra of BiFeO$_3$ ceramics reveal most of the spin excitations previously
observed in  single crystals.\cite{Talbayev11,Nagel13} Also,
at 5\,K, a pair of IR-active modes near 55\cm{} were observed, which
corresponds to spin excitations theoretically predicted
by de Sousa \textit{et al.}\cite{deSousa08} and Fishman.\cite{Fishman13b} This observation confirms the
particular form of the Hamiltonian suggested by Fishman for the
explanation of the magnetic interactions in BiFeO$_3$.

\begin{acknowledgments}
This work was supported by the Czech Science Foundation (Projects P204/12/1163 and 15-08389S), M\v{S}MT Project LH13048 and by European Union funding under the 7th Framework
Programme (Project NOTEDEV). The work at Cornell was supported by the National Science Foundation through the Penn State Center for Nanoscale Science, DMR-1420620. The work at Chatenay Malabry has been supported by the French ANR program NOMILOPS (ANR-11-BS10-016-02) project. X. F. Bai also wish to thank the China Scholarship Council (CSC) for funding his stay in France.
\end{acknowledgments}

\bibliographystyle{apsrev}

\end{document}